# Experimental Characterization of High-Amplitude Fluid-Structure-Interaction of a Flexible Hydrofoil at High Reynolds Number

Brian R. Elbing[1], Steven D. Young[2], Brent A. Craven[3], Robert L. Campbell[2], Michael L. Jonson[2], Robert F. Kunz[4] & Kevin L. Koudela[2]

[1]Mechanical & Aerospace Engineering, Oklahoma State University, Stillwater, OK 74078
[2]Applied Research Laboratory, Pennsylvania State University, State College, PA 16802
[3]Division of Applied Mechanics, U.S. Food and Drug Administration, Silver Spring, MD 20993
[4]Mechanical & Nuclear Engineering, Pennsylvania State University, State College, PA 16802

Keywords: fluid-structure-interactions, cylinder flow, time-resolved PIV

## Abstract

A fluid-structure-interaction (FSI) experiment was designed and executed with a focus on producing low-frequency (~10 Hz), high-amplitude (±3.5% of the span) fin motion. This was achieved by placing a backward facing swept fin at −9.6° angle-of-attack within the wake of a roughened cylinder. Test section speeds between 2.5 and 3.6 m/s produced cylinder diameter based Reynolds numbers between 190,000 and 280,000, respectively. Detailed descriptions of the tunnel and model geometry, material/structural behavior, fluid properties and initial conditions are provided to facilitate development of FSI models. Given the initial conditions, the resulting forced fin behavior was characterized with measurements of the mean and fluctuating components of the flow upstream of the fin (i.e. within the cylinder wake), fin-tip/surface motion and the fin constraint loading. This work provides a high fidelity experimental dataset of a challenging flow that will require two-way coupling in FSI models to properly capture the resulting behavior. Thus this rich dataset can be used by modelers to identify strengths and weaknesses of various FSI modeling approaches.

# 1 Introduction

When a fluid flows over an object it imparts stresses on the structure, which if sufficiently large can cause the structure to deform. The deformed structure modifies the flow pattern and consequently alters the stresses on the structure. Fluid-structure-interaction (FSI) modeling aims at capturing this interaction between the structure and the fluid flow. Ideally the model would be two-way coupled, which means that the fluid motion can impact the structural motion and vice versa. Two-way coupling of the flow and structural response has significantly increased in interested over recent years due to advances in computer capacity, maturing of flow/structural modeling and the realization of potential applications. There is a large body of work related to small deformation, dynamic aeroelastic/hydroelastic modeling such as flutter (Clark et al, 2004; de Langre et al., 2007) and vortex-induced vibrations (Kalmbach & Breuer, 2013; Zhao et al, 2014; Sareen et al., 2018). However, there is a dearth of available FSI literature when the structure experiences large deformations. Campbell & Paterson (2011) provides a review of available literature and reports on a method developed to compute the performance of a pump with highly flexible impellers. Flexible turbomachinery is one potential application that has gained interest amongst researchers recently (Bhavsar et al., 2009; Schmitz-Rode et al., 2005; Throckmorton & Kishore, 2009; Throckmorton et al., 2008) due to the deformable structure aiding implantation for biomedical applications. In this application, the structural response and the fluid flow are strongly coupled due to the large flow-induced deformations, which requires that the fluid flow and structural response be solved simultaneously. See Heil & Hazel (2011) for a review of internal physiological flows where FSI plays an important role, and Dowell & Hall (2001) for a review of FSI modeling approaches.

The design of marine structures, control surfaces and propellers is another potential area that could greatly benefit from advancements in FSI modeling with large structural deformations. The hydrodynamic loading on a lifting body can be significant and the resulting deformations can impact the overall performance. For example, crashback (Bridges, 2004; Bridges et al., 2008) is a common event for a ship during which the fluid loading on the structure can be extreme causing potentially significant deformations that will impact the performance of the impeller. Development of computational tools that can accurately predict the maximum stresses and deflections experienced under crashback would significantly improve the overall propeller design process. Modelling of crashback events for traditional propulsors has been an open research topic for years (Davoudzadeh, 1997; Chen & Stern, 1999; Verma et al., 2012; Jang et al., 2012). This problem will gain even greater importance as the use of composite materials (Mouritz et al., 2001; Young, 2008; Herath et al., 2014; Maljaars et al., 2018) are explored as well as new innovative and complex propulsion schemes (Triantafyllou et al, 1993, 2000; Quinn et al., 2014).

However, it is extremely challenging to perform an experiment on an unsteady fluid flow that is strongly coupled with structural motions/deformation with sufficient accuracy to be relevant for FSI model development. One approach is to use a rigid fin that is oscillated in a controlled manner, but is mounted on a flexible base such that the fin can be shifted due to the flow-induced loading. This approach was successfully studied both experimentally and numerically in Ducoin et al. (2009a, 2009b). Gomes & Lienhart (2006) provided an alternative approach for experimentally studying such a problem, where a flexible membrane was mounted in a circular cylinder wake. The tip deflections and two-component of velocity were measured at a diameter based Reynolds numbers up to 500. The current study performs a similar experiment, but instead of a flexible membrane, a fin is mounted within the cylinder wake to more accurately mimic the crashback problem. In addition, the flow is not laminar, with the cylinder diameter based Reynolds number achieving up to $2.8\times10^5$. Similar to Gomes & Lienhart (2006), the current study measures two components of velocity within the flow-field and the surface deflections, but the current study also measures the constraint loading on the fin. The current work fills a void in the literature for a high fidelity, fully coupled FSI experiment where the structure experiences large deformations.

## 2 Experimental Methods

### 2.1 Test facility and configuration

The experiment was conducted in the 12-inch re-circulating water tunnel (Deutsch & Castano, 1986; Fontaine & Deutsch, 1992; Elbing et al., 2014) at the Pennsylvania State University Applied Research Laboratory. The tunnel used a 760 mm long test section with a circular cross-section that had the diameter gradually increase from 305 mm (12-inch) at the inlet to 307 mm at the outlet to minimize flow acceleration due to the boundary layer growth. The tunnel can achieve speeds to 18 m/s and operate at pressures between 20 to 414 kPa, though for the current study the pressure was held at 128 kPa. Measurements of the empty test section free-stream turbulence intensity indicate that it was below 0.3%.

The test section layout consisted of a circular cylinder mounted with its leading edge 25 mm downstream of the test section inlet and a backward-facing fin mounted within the oscillating cylinder wake (see Figure 1). Due to the close proximity of the cylinder to the test section inlet, the inlet to the contraction immediately upstream of the test could be a preferred boundary condition for modelers. The contraction upstream of the test section had an 8.9:1 area ratio, ~1.5 m long and a fifth-order polynomial profile shape (see Elbing et al., 2018). The coordinate system used throughout this study has the $x$-axis increasing in the streamwise direction with its origin at the cylinder center. The *z*-axis increases vertically upward with the origin at the fin base height, which was 11.4 mm above the bottom of the test section at the inlet due to a ramp to produce a flat surface for mounting the fin. The ramp had an elliptical shape extending from $x = 70$ mm to $x = 119$ mm. Finally, the *y*-axis completes a right-handed coordinate system with the origin at the tunnel centerline.

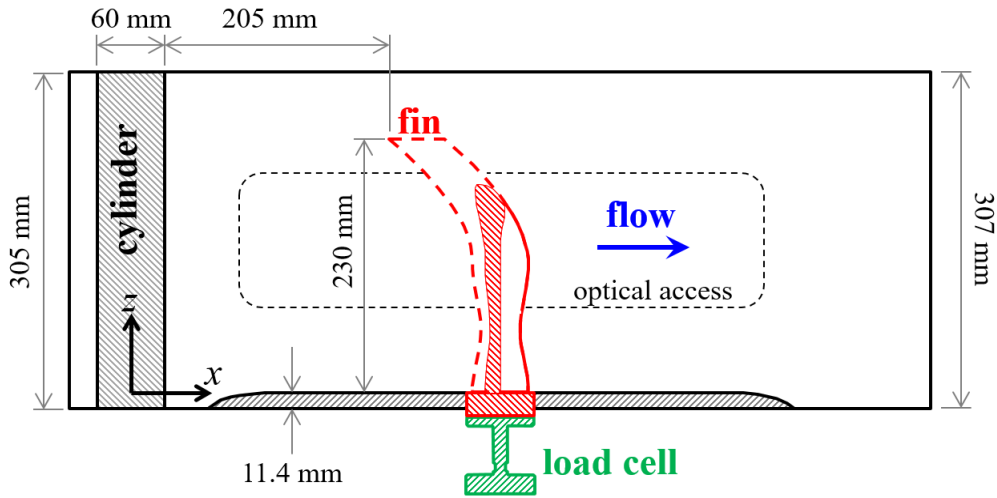

Figure 1. Cross-sectional view at the centerline ($y = 0$) of the cylinder, fin and load cell mounted in the test section. The fin was rotate 9.64° about the *z*-axis, which corresponds to an angle-of-attack for the fin of −9.64°. The optical access was 465 mm long and 120 mm tall.

## *2.2 Test model*

### *2.2.1 Cylinder*

The 60 mm diameter anodized aluminum cylinder spanned the entire test section height with its axis of rotation (centerline) located at $(x,y) = (0,0)$. The cylinder was hydraulically smooth except for the leading edge, which was roughened between $\theta = \pm 50°$ with silicon carbide grit that had an average particle size of 254 µm. Here $\theta$ is the angular position measured from the cylinder leading edge as shown in Figure 2. Preliminary data was also acquired without the roughened leading edge, and these results are included in the cylinder characterization section. The particle size and location were selected based on Nakamura & Tomonari (1982), which showed that the base coefficient and Strouhal number were independent of Reynolds number when tripped near $\theta = 50°$. As shown in Figure 2, the cylinder was instrumented with four dynamic pressure (dp) transducers (105C02, PCB) and 2 static pressure (sp) ports. The dynamic pressure transducers were located at $\theta = \pm 130°$ at the fin tip ($z = 230$ mm) and mid-span ($z = 115$ mm) heights. The static ports were at the same heights and $\theta = 180°$. A single accelerometer (303M231, PCB) was mounted inside the cylinder. Theses sensors characterized the cylinder performance, specifically the shedding frequency and relative vortex strength.

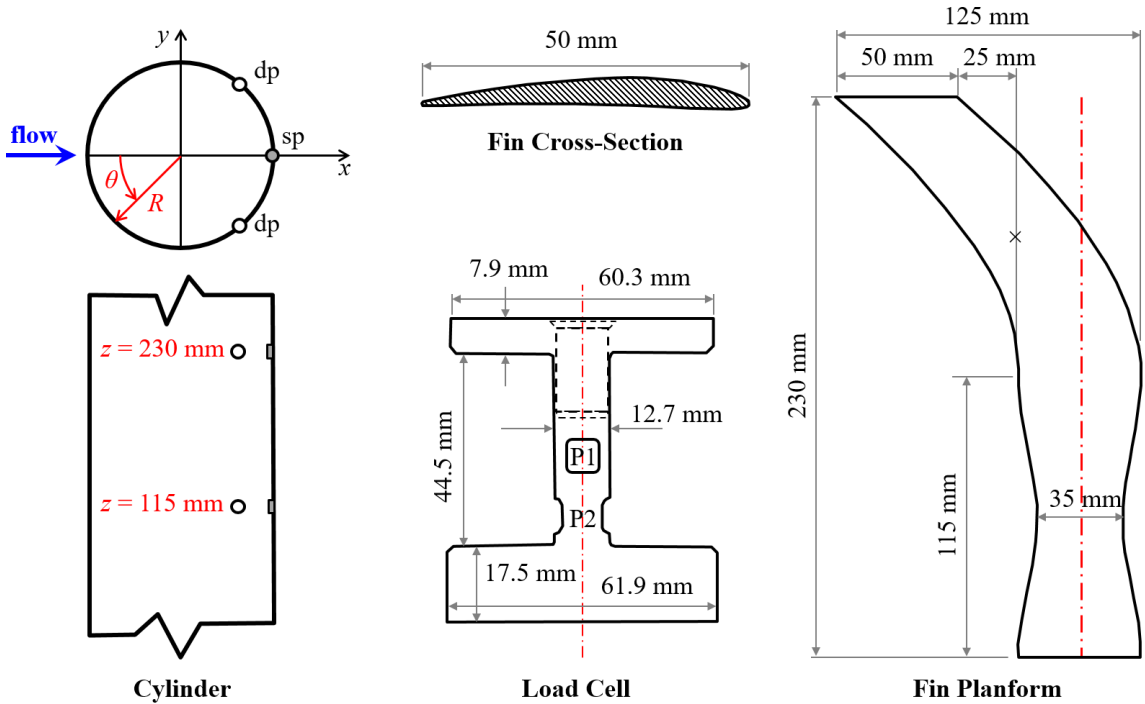

Figure 2. Sketches of the cylinder, load cell and fin (cross-section and planform views without the mounting base, as designed). For the cylinder 'dp' and 'sp' denote locations of the dynamic/fluctuating pressure and static base pressure measurements, respectively. The 7.6 mm square pockets for the (P1) drag and (P2) lift force measurements are shown. The centerlines indicate the axis that the fin/load cell can rotate about.

### *2.2.2 Fin and load cell*

The primary test model was the fin, but the load cell must also be considered since it requires the fin base to move freely for the constraint load measurements. Typically loading measurements are performed with an extremely stiff model and a relatively small gap between the model base and the surrounding structure/housing (e.g. Elbing et al., 2011). However in the current study, the model was intentionally not rigid to produce the desired large fluid-induced motions. Furthermore, the load cell flexure was weakened to improve sensitivity, which results in the load cell making a significant contribution to the total fin motion. Thus the fin and load cell (schematically shown in Figure 1 and Figure 2) assembly should be considered the test model with the base of the load cell as the fixed surface. The fin (including the 23 mm thick mounting base) was machined from a single piece of Inconel 718 with a chord length of ~50 mm, span of 230 mm and was swept forward 75 mm. The first 115 mm of span from the fin base was straight, but necked down with a constant 8% (design) maximum thickness from the chord length of 50 mm to a minimum of 35 mm at 57.5 mm from the fin base. The cross section, including the necked down region, was a NACA 4408 profile. To simulate a crashback type event, the fin was swept into the flow (opposite typical operation) and the angle-of-attack ($\alpha$) was fixed at -9.64°. Laser scanning of the as-built fin showed that the span was as designed (230 mm), the maximum chord length was slightly below 50 mm and there was a minor shift (~0.9 mm) in the *y*-direction from the tunnel centerline. The hydrodynamic leading and trailing edges at several spanwise locations as well as the corresponding chord length and angle-of-attack are provided in Table 1.

The load cell design is discussed in detail in §2.3, but here it is noted that it was fabricated from beryllium copper with the basic dimensions shown in Figure 2. The as-built fin and load cell behavior was characterized with static loading and modal impact tests. The static testing was performed by applying a point load at 75% span ($z = 172$ mm) and 62% chord (denoted by '×' in Figure 2) to the fin either with or without the load cell attached. For each configuration there were three load and unload cycles (two gradual ramps and static loading at fixed amounts). These results are shown in Figure 3 with the load point deflection converted to fin tip deflection ($\delta_t$). The fin only and fin with load cell curves have slopes of 0.104 mm/N and 0.190 mm/N, respectively.

| Span (%) | Leading Edge $x$ (mm) | Leading Edge $y$ (mm) | Trailing Edge $x$ (mm) | Trailing Edge $y$ (mm) | Chord (mm) | Angle-of-Attack (deg) |
|---|---|---|---|---|---|---|
| 1.5 | 310.2 | -5.2 | 359.2 | 3.0 | 49.7 | -9.59 |
| 11 | 313.4 | -4.7 | 356.1 | 2.7 | 43.3 | -9.81 |
| 21 | 317.1 | -3.9 | 352.4 | 2.0 | 35.7 | -9.46 |
| 31 | 316.3 | -4.0 | 353.1 | 2.4 | 37.3 | -9.77 |
| 41 | 312.1 | -4.6 | 357.1 | 3.2 | 45.7 | -9.83 |
| 51 | 309.8 | -4.9 | 359.0 | 3.5 | 49.9 | -9.71 |
| 61 | 305.4 | -5.4 | 354.4 | 3.0 | 49.8 | -9.68 |
| 70 | 295.8 | -6.7 | 344.7 | 1.5 | 49.7 | -9.51 |
| 78 | 283.0 | -8.8 | 331.8 | -0.5 | 49.5 | -9.66 |
| 86 | 268.2 | -11.3 | 316.9 | -3.0 | 49.4 | -9.73 |
| 93 | 252.1 | -13.8 | 300.7 | -5.5 | 49.2 | -9.69 |
| 100 | 235.4 | -16.3 | 283.6 | -8.5 | 48.7 | -9.24 |
| **AVG** | | | | | | **-9.64** |

Table 1. Laser scanning results from the as-built fin, including the locations of the leading and trailing edges, chord length and the corresponding angle-of-attack when installed in the tunnel.

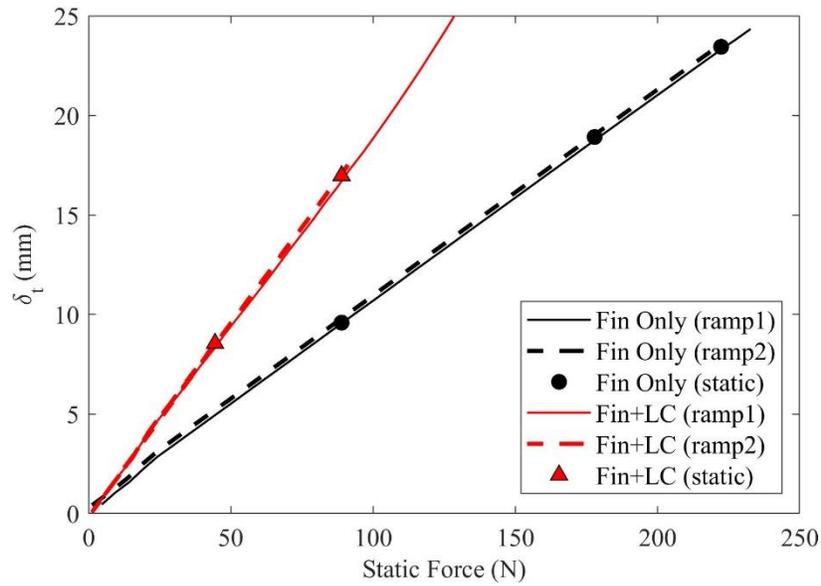

Figure 3. Fin tip deflection ($\delta_t$) from static loading of the fin only or the fin mounted on the load cell (LC). Each configuration tested two gradual ramps and a static loading.

| Mode Number | Description | Frequency (Hz) In-Air | Frequency (Hz) In-Water | Loss Factor In-Air | Loss Factor In-Water |
|---|---|---|---|---|---|
| 1 | First cross-stream bending | 32.7 | 23.2 | 0.0025 | 0.0112 |
| 2 | First streamwise bending | 44.7 | 43.9 | 0.0052 | 0.0091 |
| 3 | Second cross-stream bending | 132.0 | 102.7 | 0.0044 | 0.0069 |
| 4 | First torsion | 212.9 | 185.4 | 0.0154 | 0.0092 |
| 5 | Third cross-stream bending | 318.1 | 285.1 | 0.0017 | 0.0049 |
| 6 | Second streamwise bending | 371.0 | 341.2 | 0.0356 | 0.0587 |
| 7 | Second torsion | 586.9 | 462.4 | 0.0123 | 0.0205 |

Table 2. Frequency and loss factors for in-air and in-water modal testing of the fin and load cell assembly mounted in the water tunnel. Provided are results for the first seven mode numbers.

In-air and in-water modal impact tests were performed to experimentally determine mode shapes and loss factors of the fin mounted on the load cell. These modal tests were conducted using a roving hammer and accelerometers while the fin was installed in the water tunnel. The hit points were arranged in a 9 (spanwise) × 3 (chord-wise) grid that included 10, 25, 40, 50, 60, 70, 80, 90 and 98% span and 10, 50, and 90% chord. Two accelerometers were mounted on the opposite side of the fin from the hit points. One accelerometer was positioned at 60% span near the trailing edge and the other at 40% span near the leading edge with both oriented to measure the cross-stream ($y$) acceleration. A third accelerometer was fixed at the fin base and measured the streamwise acceleration. An additional five (10, 40, 60, 80 and 98% span) streamwise oriented hits were made on the fin trailing edge (approximately normal to the surface). A final hit was made on the fin base block. The response to these hits were sampled at 3200 Hz for a duration of 5.12 seconds, which gave a usable bandwidth of 1250 Hz. The acquired data were processed using a singular value decomposition approach combined with rational fraction polynomial curve fitting to identify mode shapes, resonance frequencies and loss factors (Fahnline, Campbell & Hambric, 2004). The frequency and loss factors of the first seven modes for the in-air and in-water cases are provided in Table 2. All seven resonant frequencies were lower in-water than in-air, but the loss factors were higher in-water than in-air for all but the first torsion mode.

## *2.3 Instrumentation*

### *2.3.1 Flow-field diagnostics*

High-speed particle image velocimetry (PIV) was acquired in double-frame, double-pulse mode at 1 kHz. The $x$- and $y$-velocity components were acquired in a single plane located at $z = 188$ mm. The image plane was illuminated with a light sheet formed from the beam (4.5 mm diameter, 527 nm wave length) of a high-speed pulsed Nd:YLF diode pumped laser (DM50-527, Photonics Industries). The laser sheet was formed with a -10 mm focal length cylindrical lens and focused with an adjustable focus lens (KL-S0-277, LaVision). The tunnel was flooded with 18 μm diameter hollow glass spheres (iM30K Glass Microspheres, 3M) to scatter the light. The scattered light was imaged with a 1280×800 pixel CMOS high-speed camera (v1610, Phantom) fitted with a 28 mm lens (EF28mm f/1.8 USM, Canon). The resulting field-of-view spanned $x$ = 100 to 340 mm and $y$ = -80 to 85 mm. The spatial calibration was performed with a precision calibration target (106-10, LaVision). A minimum of 3,000 image pairs were acquired for each test condition. The images were processed using standard cross-correlation techniques (DaVis 8, LaVision) with multiple passes. The final interrogation window was 24×24 pixels with 75% overlap, which produced a vector spacing of 1.4 mm.

Laser Doppler velocimetry (LDV) was also used for a subset of conditions to characterize the operating condition. The LDV system used an argon ion laser (Innova 70C-5, Coherent) that produced pairs of blue and green beams that were coupled to a fiber optic probe head (9832, TSI) fitted with a 350 mm lens (9253-350, TSI). The four beams were focused to a single measurement volume, which had a nominal diameter of 150 μm and length of 1.5 mm based on Gaussian beam analysis (Tropea et al, 2007). Typically 20,000 velocity measurements were acquired per condition.

### 2.3.2 Fin motion detection

Throughout testing the fin motion and orientation were continuously monitored with rotary encoder and a binary optical switch (see Elbing et al., 2014). These were to ensure that the fin tip did not deflect more than 4% of the span, which was required to maintain the infinite life criteria. However, a subset of conditions acquired precise fin tip and surface motion measurements. The fin tip deflection and rotation were acquired with high-speed imaging using the PIV setup. While raw PIV images produced low resolution fin motion measurements, a subset of conditions focused solely on the fin tip motions with high-speed imaging. The fin tip deflection (translational motion) and twist (rotational motion) were determined using cross-correlation between an image and a reference (no-flow) image. Each image had a Canny edge filter applied and divided into five chord-wise segments. The deviations from the reference image for each segment were linearly fitted, which the average offset was the fin tip translation (or deflection) and the fin tip twist was related to the change in slope. This processing scheme was compared against 400 manually inspected fin tip deflections and showed excellent agreement. Note that comparison of the reference (no-flow) images acquired throughout testing showed that the unloaded fin tip position variation was less than 0.01 mm (< 1 pixel).

The fin surface velocity was measured with a laser vibrometer (PSV-400-3D, Polytec), which was integrated to estimate local surface displacements. The system used a HeNe laser (633 nm wavelength) that was sensitive to frequencies from near DC to 1 MHz. The vibrometer could scan up to 30 surface locations per second and was setup to acquire measurements at 21 locations on the fin surface. The points were laid out in a 7×3 grid with seven spanwise (40, 50, 60, 70, 80, 90 and 95% span) and 3 chord-wise (10, 50 and 90% chord) positions. Optical access prevented measurements at the fin tip (100% span). The beam was aligned such that it was perpendicular to the fin surface at the 90% chord location when the fin was at the no-flow angle-of-attack (-9.64°). Corrections for the relative surface angle at each measurement location were performed assuming purely normal (translational) surface motion, which this assumption is supported subsequently with the fin tip twist results. A second laser vibrometer (OFV-505/5000, Polytec) was fixed at 80% span, 90% chord to serve as a reference point. The signals were low pass filtered at 5 kHz and sampled at a rate of 512 Hz. The frequency resolution was set at 0.125 Hz, which required an 8 second period per ensemble average. A total of 16 complex ensembles were acquired at each measurement location using 50% overlap processing. The acquired data was recorded with the system analyzer as well as via an external data acquisition system (PXI 4496, NI). Following the analysis of Bendat & Piersol (1980) assuming a 0.9 coherence, the normalized random errors for the autospectrum, cross-spectrum, coherence, coherent output power and transfer function were 0.250, 0.264, 0.037, 0.276 and 0.0589, respectively.

### *2.3.3 Fin constraint loading*

The fin constraint loading was measured with a custom 4-component load cell, which had an instrumented cylindrical flexure (see Figure 2 for schematic). The instrumented section had an outer diameter of 12.7 mm, was 44.5 mm long and had four 7.62 mm square pockets (P1 & P2 in Figure 2) cut into the shaft to produce two thin membranes on the shaft centerline for the lift and drag measurements. The lift and drag membranes were rotated 90° relative to each other and centered 45.4 and 31.9 mm below the fin attachment location, respectively. A 12.0 mm diameter, 22.2 mm deep hole centered at the fin attachment created a 0.71 mm thick thin-walled section to facilitate the torque and drag-moment measurements. Each cell consisted of a Wheatstone bridge of strain gauges (lift, drag, torque: sk-06-062TW-350, Vishay; *y*-moment: TK-06-092P-10C/DP, Vishay) affixed to the flexure. All bridges were excited with a 10 V DC signal, and the output was sampled at 1 kHz. With the fin at $\alpha = 0°$, the load cell measured the lift force ($\mathcal{L}$), drag force ($\mathcal{D}$), *y*-axis moment (drag-moment; $M_D$) and the *z*-axis moment (torque; $T_q$), which were designed for maximum loads of 110 N, 22 N, 5.0 Nm and 5.5 Nm, respectively. Since the test angle-of-attack was −9.64°, the load cell output was corrected to determine the true lift and drag forces.

A calibration matrix was produced from 356 measurements using a single load (53.4 N) hung from 3 spanwise locations, 2 torque moment arms (0 or 25.3 mm) and fin rotations between 0° (forced aligned with the chord) and 315°. This benchtop calibration was confirmed with an *in-situ* calibration with forces applied at various fin locations. The instrument's random and bias errors were quantified using the residual loads, difference between back-calculated loads and known applied calibration loads. The mean residual loads are provided in Table 3, which given the nominal test range the uncertainty in the lift force and drag-moment were ~5% while the drag and torque were between 7% and 10%. The residual loads do not account for additional bias errors from misalignment between the floating element (fin base) and the surrounding tunnel wall as well as flow through the gap. The misalignment was ~100 μm, which experimental results (Allen, 1976; Klewicki, 2007) indicate that for the current gap (0.3 mm) and nominal boundary layer thickness (~10 mm) the uncertainty is on same order magnitude as the skin-friction generated on the fin base. This is considered negligible since the base skin-friction was much smaller than the fin loading (both in surface area and force magnitude). Additional details on wiring diagram, calibration procedures and uncertainty analysis can be found in Elbing et al. (2014).

| Component | Mean Residual Load | Full Scale | Typical Uncertainty |
|---|---|---|---|
| Lift Force ($\mathcal{L}$) | ±0.3% f.s. | 110 N | ±0.3 N |
| Drag Force ($\mathcal{D}$) | ±2.0% f.s. | 20 N | ±0.4 N |
| Torque ($T_q$) | ±1.0% f.s. | 5.4 Nm | ±0.05 Nm |
| Drag Moment ($M_D$) | ±0.5% f.s. | 4.7 Nm | ±0.02 Nm |

Table 3. Mean residual loads as a percentage of full scale (f.s.) from the load cell calibration and the resulting uncertainty range for each measured component.

### *2.3.4 Tunnel and cylinder monitoring*

Throughout testing the water temperature, tunnel pressure (total and static) and the tunnel impeller frequency were recorded at 1 kHz. The temperature was measured with a resistance temperature detector (RTD) sensor (911PL, Stow Laboratories). The total pressure was measured with a Kiel probe mounted upstream of the test section contraction. Static pressure was measured at the total pressure location as well as along the test section length ($x$ = 0, 75, 175, 280, 385 and 490 mm). The local free-stream speed was determined from the difference between the total and static pressure, which was confirmed with PIV and/or LDV measurements.

## *2.4 Test conditions*

While the current manuscript focuses on the fin mounted downstream of a roughened cylinder, data were also acquired for three other test section configurations; (i) empty test section, (ii) smooth cylinder only and (iii) rough cylinder only. Full details of these other tests are provided in Elbing et al. (2014), but some data from these other configurations are included in the current manuscript to assist in interpretation of results as well as provide additional validation data for modelers. During the forced fin testing the fin angle-of-attack was fixed at -9.64°, and the tunnel impeller frequency was varied between 50 and 84 rpm. The average water temperature was 20.5 ± 0.5 °C, which has a corresponding water density ($\rho$) and kinematic viscosity ($\nu$) of 998 kg/m$^3$ and 9.9×10$^{-7}$ m$^2$/s, respectively.

## 3 Flow Inlet Characterization

Due to the close proximity of the cylinder to the test section inlet, it was not possible to directly measure the upstream velocity to prescribe a mass flowrate for modelers. Table 4 provides three measures of free-stream downstream of the cylinder for each tunnel impeller frequency ($f_{imp}$). Here $U_{sp}$ is the average free-stream determined from static pressure taps spanning the fin location ($x$ = 175 to 490 mm), $U_C$ is the centerline ($y$ = 0) streamwise velocity one chord length upstream of the fin tip ($x$ = 185 mm) and $U_\infty$ is direct measurement (PIV) of the free-stream a given downstream position. The variation between the PIV and static pressure determined free-stream speeds was within ~2%. Given these measurements modelers can adjust their inlet condition to match the downstream speeds. To assist with the initial guess, the nominal mass flowrate was estimated from the wake profiles. The mean wake velocity profiles scaled like a self-similar plane wake (Pope, 2000) are given in Figure 4, where $U$ is the local mean streamwise velocity, $U_S(= U_\infty - U(x, y = 0))$ is the velocity deficit and $y_{0.5}$ is the wake half-width defined as $U(x, \pm y_{0.5}) = U_\infty - 0.5 U_S(x)$. The current results approach a self-similar constant turbulent viscosity plane wake, $(U_\infty - U)/U_S = \exp(-0.693\, y/y_{0.5}(x))$, with increasing downstream distance. An estimate of the mass flowrate $(\dot{m})$ can be determined by integrating the wake profiles assuming a constant free-stream speed outside of the wake (i.e. ignoring wall boundary layers). The profile scaling parameters and resulting mass flowrates are provided in Table 4.

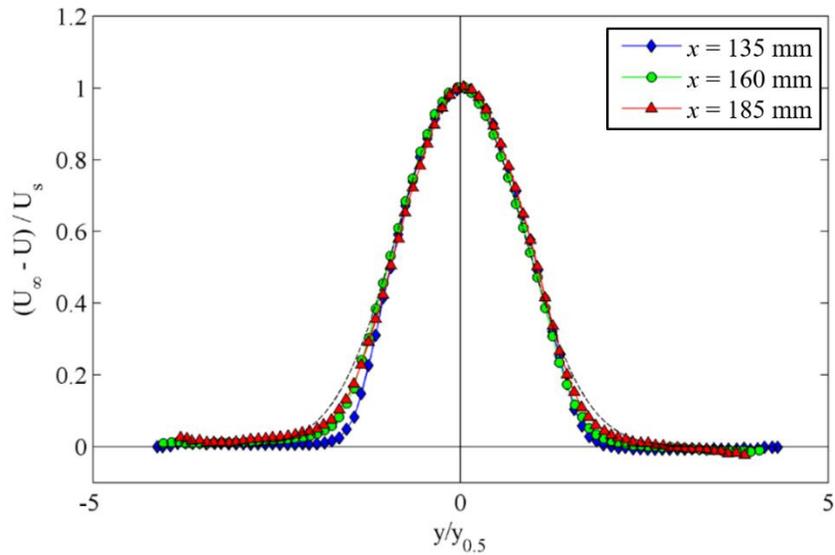

Figure 4. Normalized average wake velocity deficit profiles at three streamwise locations. The dashed line corresponds to a self-similar plane wake with constant turbulent viscosity (Pope, 2000).

| $f_{imp}$ (Hz) | $U_{sp}$ (m s$^{-1}$) | $U_C$ (m s$^{-1}$) | $x$ (mm) | $U_\infty$ (m s$^{-1}$) | $U_S$ (m s$^{-1}$) | $y_{0.5}$ (mm) | $\dot{m}$ (kg/s) |
|---|---|---|---|---|---|---|---|
| 58.3 | 2.51 | 1.19 | 135 | 2.59 | 2.51 | 21.1 | 167 |
|  |  |  | 160 | 2.48 | 1.79 | 22.2 | 166 |
|  |  |  | 185 | 2.42 | 1.25 | 25.8 | 165 |
| 69.6 | 2.99 | 1.13 | 135 | 3.21 | 3.29 | 22.7 | 202 |
|  |  |  | 160 | 3.09 | 2.56 | 22.3 | 202 |
|  |  |  | 185 | 3.00 | 1.89 | 23.4 | 202 |
| 80.1 | 3.44 | 1.11 | 135 | 3.66 | 3.91 | 23.9 | 225 |
|  |  |  | 160 | 3.48 | 3.10 | 22.3 | 224 |
|  |  |  | 185 | 3.36 | 2.27 | 22.3 | 226 |
| 84.2 | 3.62 | 1.57 | 135 | 3.89 | 3.80 | 23.2 | 246 |
|  |  |  | 160 | 3.77 | 2.89 | 23.6 | 247 |
|  |  |  | 185 | 3.69 | 2.13 | 25.1 | 249 |

Table 4. Tunnel impeller frequency ($f_{imp}$), corresponding average free-stream speeds and scaling parameters for the mean wake deficit profiles for each streamwise position as well as the resulting estimate of the mass flowrate ($\dot{m}$).

# 4 Cylinder Characterization

## 4.1 Base pressure

The flow around the cylinder in three configurations (smooth cylinder, rough cylinder and rough cylinder with fin installed) were characterized with the base static and dynamic pressures. Since the drag on a cylinder is directly related to the base static pressure, it is common to use the base pressure coefficient ($C_{pb} = \Delta P_b / 0.5\rho U_\infty^2$) to characterize the flow, where $\Delta P_b$ is the difference between the base pressure ($\theta = 180°$) and the local static pressure. The data for all three configurations at the fin mid-span ($z = 115$ mm) and tip ($z = 230$ mm) heights are provided in Figure 5 as a function of the diameter based Reynolds number ($Re_D = U_\infty D/\nu$). Note that the freestream speed was determined from the downstream measurements and accounting for blockage effects (multiplied by 1.334). A representative smooth cylinder curve (Shih *et al*., 1993) is also provided for reference. The smooth cylinder results follow the trend of the smooth cylinder curve with scatter that is consistent with historical data (Bearman, 1969; Nakamura & Tomonari, 1982; Shih *et al*., 1993). These results also show that the smooth cylinder was within the transitional range, which motivated the roughening of the leading edge. The rough cylinder results closely follow the data used to size and position the roughness (Nakamura & Tomonari, 1982), which has a weak Reynolds number dependence. The current results show minimal variation between spanwise locations with and without the fin installed. However, there is a small but measureable shift in the cylinder performance with and without the fin installed within the cylinder wake. This slight shift was likely due to the oscillating fin impacting the free-stream speed measurement, which has an impact on the static pressure (not the actual cylinder flow-field). In spite of the variation with and without the fin as well as the increased scatter, it is apparent that with the fin installed there was minimal spanwise variation and the behavior is nearly Reynolds number independent over the operating range.

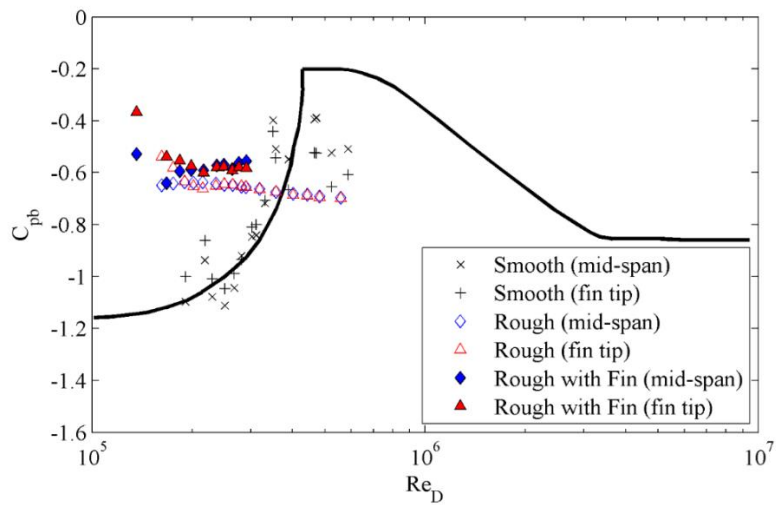

Figure 5. Base pressure coefficient versus cylinder diameter based Reynolds number. A representative smooth cylinder curve (Shih *et al*., 1993) is included for reference.

## *4.2 Shedding frequency*

The vortex shedding frequency from the cylinder was determined from spectral analysis of the dynamic pressure on the aft side of the cylinder. The shedding frequency and root-mean-squared (RMS) levels on the smooth cylinder increased with increasing speed until ~3 m/s ($Re_D = 1.8 \times 10^5$), at which point the peak began to dissipate and completely disappeared above 4 m/s ($Re_D = 2.4 \times 10^5$). This was due to the laminar-to-turbulent transition of the cylinder boundary layer, which results in the wake becoming incoherent due to the moving separation point. Conversely, the rough cylinder showed increasing frequency and RMS levels up to the max speed of 6 m/s ($Re_D = 3.6 \times 10^5$). The rough cylinder had three spectral peaks in addition to the shedding frequency peak. Representative rough cylinder power density spectra ($\Phi_{dp}$) at $U_{sp} = 2.99$ m/s without the fin installed is provided in Figure 6 along with a reference no-flow spectra. Spectral peaks in the dynamic pressure are observed at 13 (shedding frequency), 16, 17.1 and 18.6 Hz. The shedding frequency was determined using the pairs on each side of the cylinder to estimate the fluctuating lift and drag spectra, which the drag spectra is double the shedding frequency in the lift force. This also revealed that the higher frequency peaks were only associated with the drag force. This indicates that the most likely cause for the complex drag spectrum was spanwise variation since LDV measurements of the tunnel speed confirmed that it was stable and repeatable. Fortunately, the fin loading is primarily dependent on the lift component, which does not exhibit the complex frequency content.

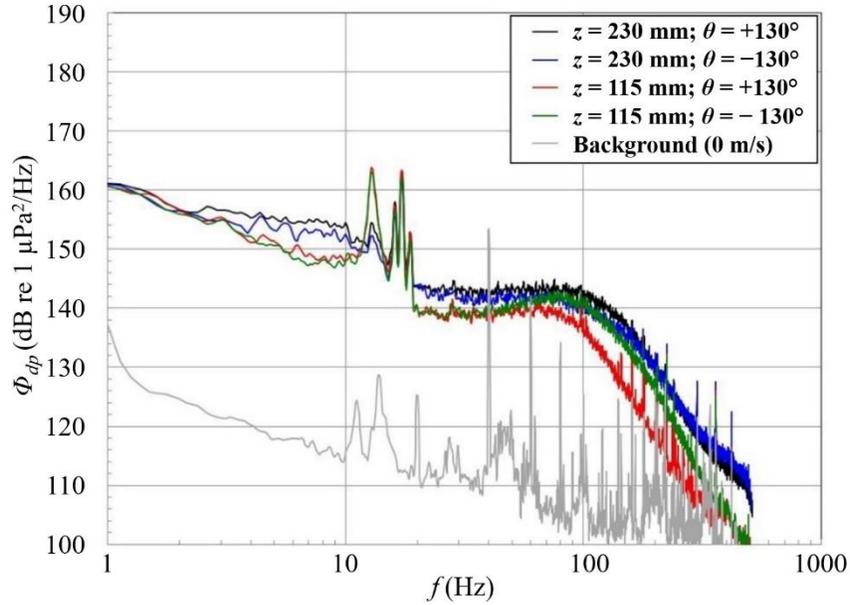

Figure 6. Pressure spectra from the rough cylinder without the fin at $U_{sp} = 2.99$ m/s. Spectra are shown from all four sensors located at $z = 115$ or 230 mm and $\theta = \pm 130°$. Also included is the electrical background spectrum (0 m/s) from a single sensor.

All three configurations had excellent spanwise coherence on both sides of the cylinder at the shedding frequency. The smooth cylinder had the highest coherence (0.9) due to the laminar boundary layer on the cylinder. The influence of the fin on the cylinder performance was examined by comparing spectra with and without the fin installed. In spite of the slight variation in the base pressure, comparison of the rough cylinder fluctuating pressure spectra with and without the fin shows that there was no significant variation in peak RMS levels at the shedding frequency. This supports the previous conjecture that the base pressure coefficient variation with and without the fin was the product of the fin's influence on the velocity measurement and not the cylinder performance.

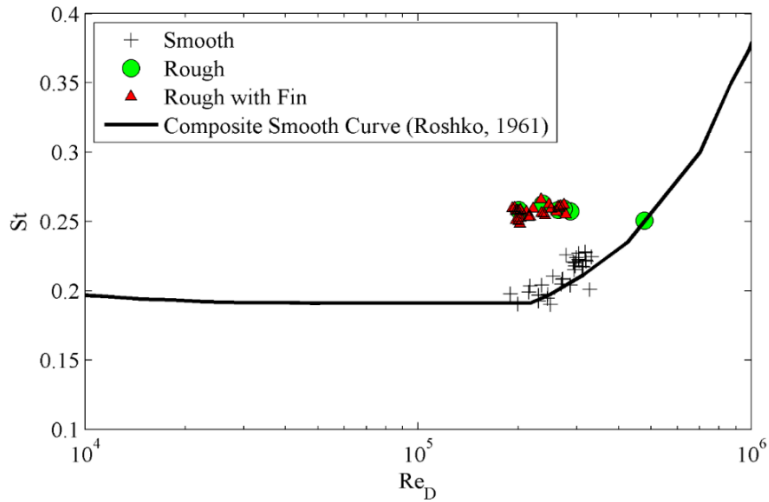

Figure 7. Strouhal scaling of the shedding frequency from the smooth, rough and rough with the fin installed cylinder configurations plotted versus the Reynolds number. Included for comparison is a representative composite curve for smooth cylinders (Roshko, 1961).

The cylinder vortex shedding frequency ($f_s$) results are provided for all three configurations in Figure 7 with the Strouhal number ($St = f_s D/U_\infty$) plotted versus $Re_D$. A composite curve from Roshko (1961) for smooth cylinders is included for reference. The current smooth results are in good agreement with the composite curve, but no coherent shedding was observed for $Re_D \geq 3.3 \times 10^5$. Conversely, the rough cylinder had a nearly constant Strouhal number (0.26) over the range of Reynolds numbers tested. Nakamura & Tomanari (1982) had a slightly higher Strouhal number, but both exhibit nearly constant value. Finally, there was negligible difference observed with and without the fin installed. The bias error was quantified by examining the variation between independent measurements (dynamic pressure, cylinder accelerometer and LDV wake measurements) while the random error determined from comparison between repeated measurements from a single test condition. The bias and random errors were ±1.2% and ±2% with a 95% confidence level, which gives a shedding frequency accuracy of ±2.3%.

# 5   Forced Fin Results

## 5.1  Flow-field (cylinder wake)

### 5.1.1  Mean and fluctuating velocity profiles

The flow-field upstream of the fin was characterized with the high-speed (1 kHz) PIV, which measured the streamwise ($u$) and the cross-stream ($v$) velocity within a 2D plane at 82% span ($z$ = 188 mm). A representative video of the instantaneous vector fields superimposed on background images at $U_{sp}$ = 3.62 m/s are provided in supplementary material, Movie 1. The cylinder is located ~100 mm upstream (to the left), the fin motion can be observed on the right side and the lack of vectors in the lower right corner was due to the shadow cast by the fin. The video shows the vortex structures shed from the cylinder and convect downstream until they impact the fin resulting in the fin movement. From the instantaneous vector fields, the mean and fluctuating velocity profiles for both the streamwise ($x$) and cross-stream ($y$) directions were extracted. Figure 8 provides the average streamwise ($U$), fluctuating streamwise ($u'$), average cross-stream ($V$) and fluctuating cross-stream ($v'$) velocity profiles scaled with $U_{sp}$ and cylinder diameter $D$ from three streamwise locations ($x/D$ = 2.25, 2.67, 3.08).

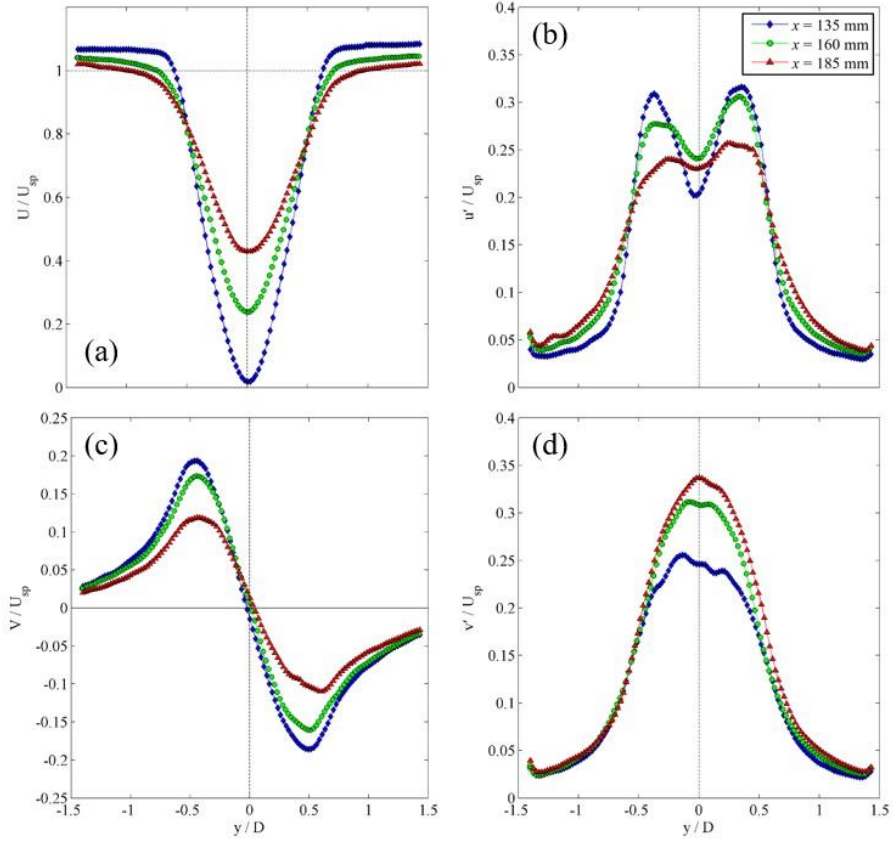

Figure 8. The mean (a,c) and fluctuating (b,d) velocity profiles scaled with the tunnel free-stream speed ($U_{sp}$ = 3.62 m/s) plotted versus the cross-stream position scaled with the cylinder diameter (*D*). The top and bottom row are the streamwise and cross-stream components, respectively.

The upstream blockage created by the cylinder causes the mean streamwise velocity outside of the wake to exceed $U_{sp}$ in the near-wake region, which this effect decreases with increasing downstream distance. As expected, the mean and fluctuating streamwise velocity profiles show that with increasing downstream distance the wake deficit decreases and the wake spreads. The streamwise fluctuating profiles have a bimodal distribution with peaks located at nominally ±0.4*D*, which also decreases with increasing downstream distance. However, due to the spreading wake the fluctuating velocity increases with increasing downstream distance for $|y/D| > 0.6$. The mean cross-stream velocity profiles are approximately zero at the centerline and symmetric in magnitude about the centerline. The peaks occur near $y/D = \pm 0.5$, though the positive peak appears to be shifted slightly towards the centerline. Since the range of positions shown are beyond the recirculation region ($x/D \sim 2.2$), the peaks decrease with increasing downstream distance. The centerline fluctuating cross-stream velocity increases with increasing downstream distance, but this trends ceases beyond $x = 185$ mm. These results are not shown as the goal is to show the cylinder wake behavior prior to impact the fin.

An alternative approach to assess the mean velocity field is to examine the average flow-field effective angle-of-attack relative to the streamwise direction, $\alpha_{\text{eff}} = \tan^{-1}(V/U)$. Figure 9 shows the effective angle-of-attack profiles from the same streamwise profiles shown in Figure 8. These profiles are symmetric in magnitude about the tunnel centerline with the peak magnitudes decreasing and moving away from the centerline with increasing downstream distance. This is due to the increasing $U$ along the centerline, which is zero at the end of the recirculation zone. In fact, the width of the recirculation zone becomes readily identified when plotting $\alpha_{\text{eff}}$ within that region because two asymptotes appear at the edges. While only a single speed is shown, there was excellent agreement independent of the test speed. If profiles were shown farther downstream they approach a constant zero value. This does not indicate that there were not large fluctuations, but that the velocity fluctuations across the entire wake fluctuate about zero degrees.

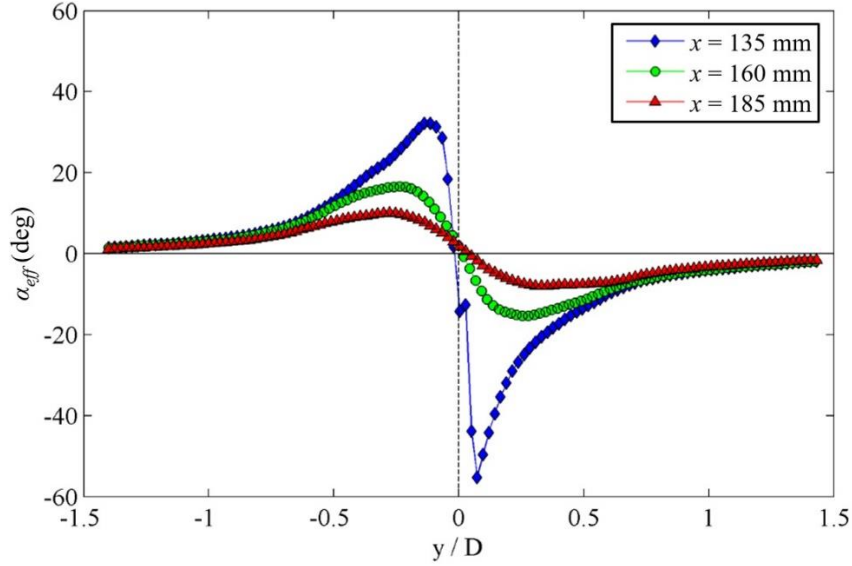

Figure 9. The effective angle-of-attack, $\alpha_{\text{eff}} = \tan^{-1}(V/U)$, plotted versus the cross-stream position scaled with the cylinder diameter for the three streamwise positions shown in Figure 8.

### *5.1.2 Spectral results*

The spatial correlations and wavenumber spectra were computed from the vector fields to examine the spatial distribution of the fluctuating content in the velocity field. The spatial autocorrelation for a homogeneous velocity component *u* is defined as $R_{uu}(\xi,\eta) = E[u(x,y)u(x+\xi,y+\eta)]$, where *E*[--] is the expected value and $\xi$ and $\eta$ are relative distances in the *x* and *y* directions, respectively. The spatial autocorrelation is even (symmetric), thus all spatial autocorrelations presented here are shown for positive separation distances only. Contour plots of the spatial autocorrelation distribution for *u* and *v* are included in Elbing et al. (2014) with the general shape similar between speeds and the peak correlations increasing with velocity. Figure 10 provides the autocorrelation for *u* and *v* for a range of streamwise (*Δx*) and cross-stream (*Δy*) separation distances. The autocorrelations have been normalized by the variance ($\sigma_u^2$ or $\sigma_v^2$), which was 0.02 m²/s² for the cross-stream velocity (*v*) and the streamwise velocity (*u*) was 0.16, 0.2 and 0.35 m²/s² for $U_{sp}$ = 2.51, 2.99 and 3.62 m/s, respectively. The streamwise variance was proportional to the tunnel speed squared, and all the profiles have similar trends for a given speed and direction. The variation for both velocity components in the streamwise direction (*x*) appear to be oscillating, which is typical for vortex shedding within a cylinder wake. As expected, the oscillating behavior is not observed in the cross-stream direction since the wake is not periodic in that direction.

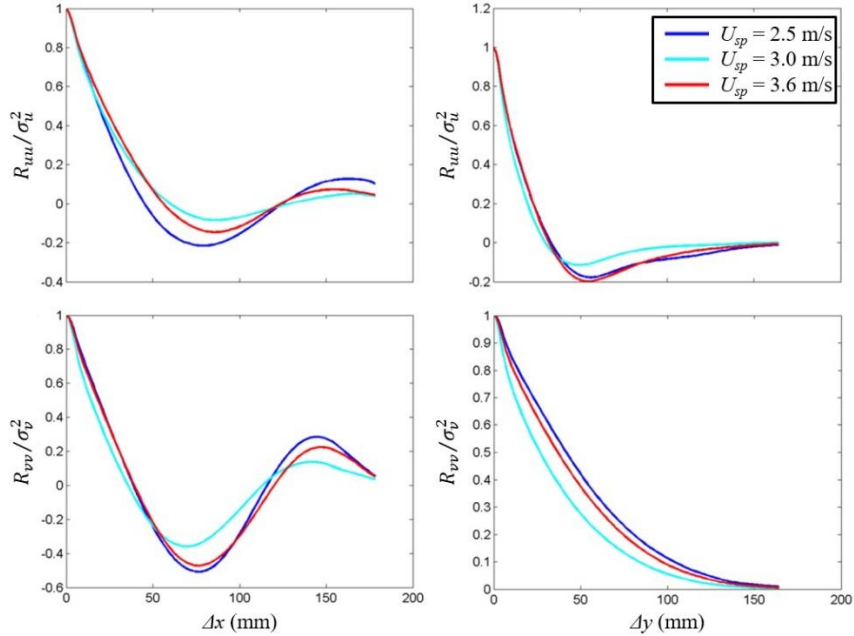

Figure 10. Comparison of the velocity spatial correlations along the *x*- and *y*-axis for the *x*- and *y*-components of velocity.

The wavenumber (*k*) spectrum is the Fourier transform of the autocorrelation in both *x* and *y* directions,

$$S_{uu}(k_x, k_y) = \int_{-\infty}^{\infty} \int_{-\infty}^{\infty} R_{uu}(\xi, \eta) \exp[-i(k_x \xi + k_y \eta)] d\xi d\eta.$$

Since the spatial autocorrelation is even along both axes, the corresponding autospectrum is real and even (symmetric). Thus, the wavenumber autospectra are only shown for positive wavenumbers ($k_x$, $k_y$). The indirect fast Fourier transform (FFT) approach (Bendat & Piersol, 1986) was used to compute these wavenumber spectra. The range was extended by padding the velocity with zeroes and a Hanning window was applied to the spatial data. The FFT of the extended velocity data was used to estimate the autospectral density, and then the inverse FFT of the autospectral density gives the autocorrelation. Assuming a coherence of 1.00, the normalized random error for the autospectrum was 0.154 or ~0.6 dB. Contour plots of the spatial distribution of $S_{uu}$ and $S_{vv}$ are provided in Elbing et al. (2014), and Figure 11 provides curves extracted from the contour plots of the autospectra versus $k_x$ for various constant $k_y$ values. Comparison of these wavenumber spectra with those acquired without the fin installed (not shown) show negligible difference. This shows that at this location there was no apparent feedback mechanism between the upstream flow-field and the fin.

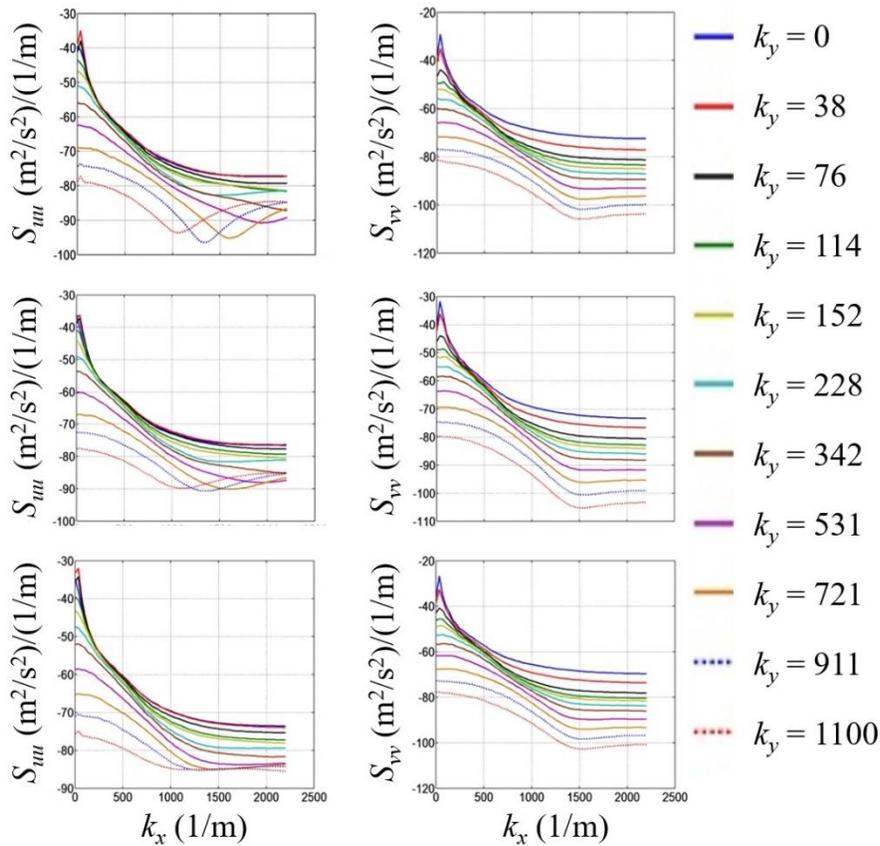

Figure 11. Wavenumber spectra for the (left) *x*- and (right) *y* velocity components at $U_{sp}$ = (top) 2.52, (middle) 2.99 and (bottom) 3.62 m/s. The bandwidth along $k_x$ is 35.1 m$^{-1}$.

## 5.2 Fin motion

The fin tip refers to the top of the fin where the maximum deflections (translational motion) and twist (rotation) occur, which was measured with high-speed imaging. The fin tip twist ($\phi_t$) results reveal a linear relationship with the tunnel speed, $\phi_t = 0.20 U_{sp} - 0.372$, between $2.5 \leq U_{sp} \leq 3.4$. In general, there was minimal twist with the maximum observed mean and RMS twist being 0.31° and 0.33°, respectively. Fin tip deflection mean ($\delta_t$) and RMS ($\delta_t'$) values scaled by the fin span ($b = 230$ mm) as a function of the chord ($c = 50$ mm) based Reynolds number ($Re_c = U_{sp} c / \nu$) are provided in Figure 12. The mean and RMS fin tip deflections were linearly fitted resulting in $\delta_t/b = -1.13 \times 10^{-7} Re_c + 0.011$ and $\delta_t'/b = 9.85 \times 10^{-8} Re_c - 0.0083$, respectively. Note that fitting these results with a power-law were equally valid, but without a large Reynolds number range it is difficult to accurately estimate the power. Also included in the figure are dashed lines that show the nominal peak deflections observed during testing, which were nominally ±2.9 times the mean deflections. Thus the peak-to-peak fin deflections were up to 5.8 times the magnitude of the mean tip deflection. The mean deflection became more negative with increasing speed due to the negative angle-of-attack. The uncertainty in the mean and RMS tip deflections are ±7% (≤ ±0.15 mm) and ±9% (≤ ±0.2 mm) of the measured values, respectively. This uncertainty was primarily due to the combination of a relatively short sample time (3 sec) and broadband frequency content observed in the fin tip motion.

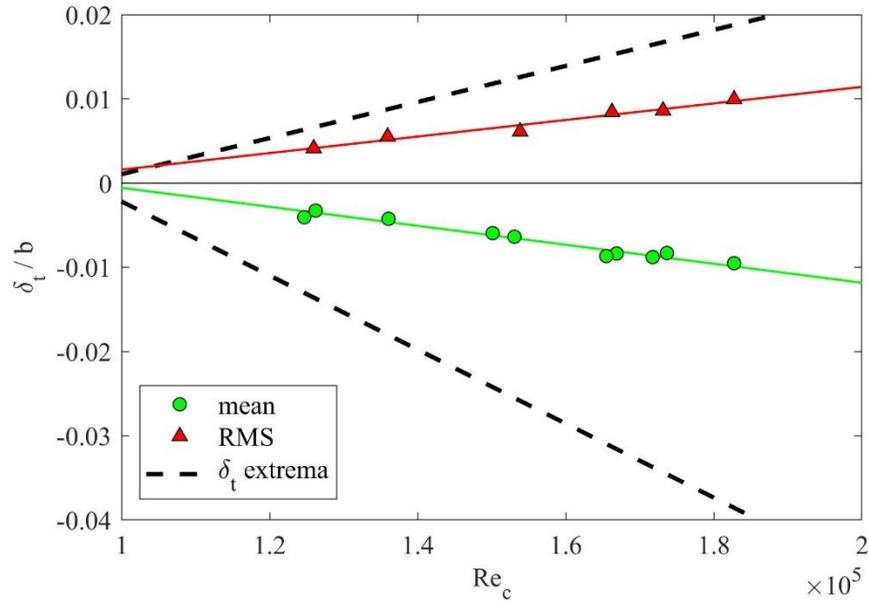

Figure 12. Fin tip deflections (mean and RMS) scaled with the fin span (230 mm) and plotted versus the chord based Reynolds number. Dashed lines correspond to the nominal range fin tip deflection extrema. Solid lines are the linear best-fit curves for the mean and RMS deflections.

Fin tip time traces reveal broadband frequency content in the fin motion but spectral density was not performed due to the limited sample period (3 sec). However, the peak in the autocorrelation shows that the dominate frequency was nominally equal to the cylinder vortex shedding frequency. Consequently, spectral density of the fin motion was examined with the surface laser vibrometer. A typical deflection RMS amplitude spectrum $\Phi_\delta$ (square root of the autospectrum level) from a single point (90% span, 50% chord) is shown in Figure 13. Displacements were determined from the integral of the surface velocity. Two peaks were observed below 50 Hz, with the first and second corresponding to the cylinder shedding frequency and first fin bending mode, respectively. The peak at ~100 Hz corresponds to the in-water second cross-stream bending mode with no apparent peak at the first streamwise bending mode (~44 Hz), which is consistent with the fact that the surface vibrometer measures the motion normal to the surface (i.e. more sensitive to cross-stream motion than streamwise). Examination of all these points allow for estimates of the relative surface motion at a given frequency (*f*). As expected given the deflection spectrum, the predominant mode near the shedding frequency is the first bending with the largest deflections occurring near the fin tip. See Elbing et al. (2014) for example contour plots. The surface deflection results are tabulated in Table 5, which includes the peak frequency ($f_{pk}$), the corresponding RMS level and the quality factor. The quality factor $\left(Q = f_{pk}/(f_U - f_L)\right)$ is a measure of the bandwidth of the peak, where $f_U$ and $f_L$ are the frequency 3-dB below the peak on the upper and lower side of the peak, respectively. In addition, the average and standard deviation of $f_{pk}$ and $Q$ are provided. These results show that $f_{pk}$ increases with speed but remains relatively stable independent of speed. The width of the peak broadens (i.e. $Q$ increases) with increasing speed and experiences larger fluctuations.

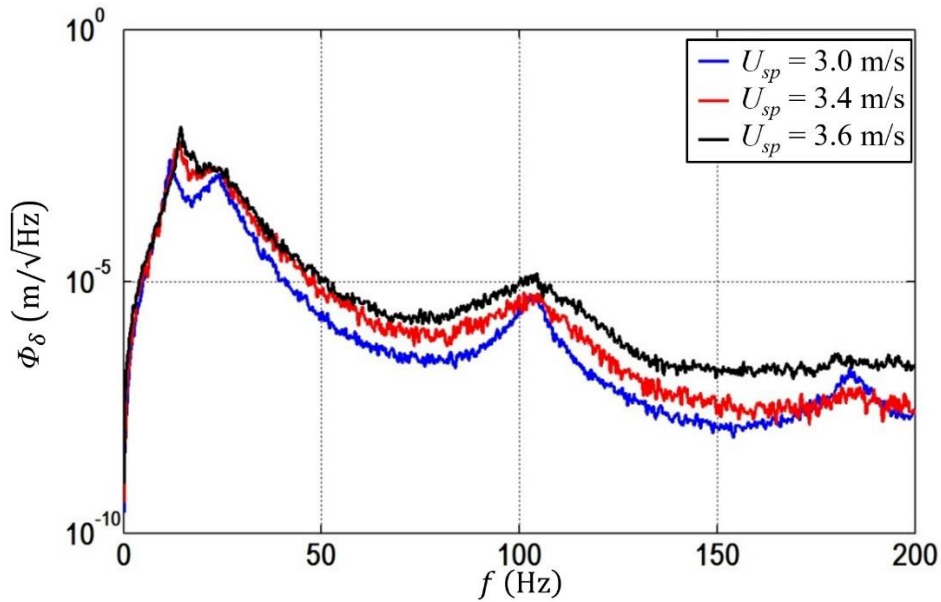

Figure 13. Fin deflection RMS spectra acquired near the fin tip (90% span) and 50% chord position.

|  |  | $U_{sp}$ = 2.99 m/s | | | $U_{sp}$ = 3.44 m/s | | | $U_{sp}$ = 3.62 m/s | | |
|---|---|---|---|---|---|---|---|---|---|---|
| Span (%) | Chord (%) | $f_{pk}$ (Hz) | Peak (RMS m/s) | Q (--) | $f_{pk}$ (Hz) | Peak (RMS m/s) | Q (--) | $f_{pk}$ (Hz) | Peak (RMS m/s) | Q (--) |
| 40 | 90 | 12.0 | 0.44 | 10.7 | 13.9 | 1.49 | 12.3 | 14.5 | 2.49 | 19.3 |
| 40 | 50 | 11.9 | 0.41 | 9.5 | 14.0 | 1.01 | 10.2 | 14.5 | 2.25 | 14.5 |
| 40 | 10 | 11.5 | 0.32 | 11.5 | 13.5 | 0.91 | 9.0 | 14.0 | 1.55 | 8.0 |
| 50 | 90 | 11.9 | 0.58 | 7.3 | 14.3 | 1.95 | 11.4 | 14.4 | 2.22 | 9.6 |
| 50 | 50 | 11.6 | 0.42 | 6.2 | 13.8 | 2.70 | 11.0 | 14.5 | 3.98 | 14.5 |
| 50 | 10 | 11.9 | 0.63 | 9.5 | 13.5 | 1.77 | 8.3 | 14.5 | 2.51 | 8.9 |
| 60 | 90 | 12.1 | 0.96 | 6.9 | 13.8 | 3.92 | 15.7 | 14.3 | 4.04 | 6.0 |
| 60 | 50 | 11.9 | 1.10 | 13.6 | 13.6 | 2.99 | 9.9 | 14.1 | 3.25 | 8.7 |
| 60 | 10 | 12.0 | 0.72 | 7.4 | 13.6 | 3.16 | 9.9 | 14.4 | 3.86 | 9.6 |
| 70 | 90 | 11.9 | 1.26 | 9.5 | 13.8 | 5.48 | 11.0 | 14.3 | 6.01 | 14.3 |
| 70 | 50 | 11.9 | 1.25 | 7.9 | 13.8 | 2.94 | 4.6 | 14.0 | 4.17 | 8.0 |
| 70 | 10 | 12.4 | 1.62 | 9.0 | 14.0 | 5.11 | 10.2 | 14.8 | 5.34 | 14.8 |
| 80 | 90 | 11.6 | 1.51 | 4.7 | 13.8 | 4.34 | 6.1 | 14.8 | 8.18 | 16.9 |
| 80 | 50 | 12.3 | 1.41 | 6.5 | 14.1 | 5.82 | 9.4 | 14.3 | 7.14 | 10.4 |
| 80 | 10 | 12.1 | 1.57 | 6.9 | 14.0 | 4.27 | 9.3 | 14.4 | 8.07 | 8.9 |
| 90 | 90 | 11.6 | 2.25 | 8.5 | 13.6 | 6.63 | 8.4 | 14.8 | 10.6 | 13.1 |
| 90 | 50 | 11.6 | 2.29 | 6.6 | 14.0 | 5.60 | 11.2 | 14.6 | 10.6 | 9.8 |
| 90 | 10 | 11.8 | 2.22 | 5.5 | 13.5 | 5.15 | 6.0 | 14.4 | 11.2 | 11.5 |
| 95 | 90 | 11.8 | 2.61 | 9.4 | 13.5 | 6.61 | 7.7 | 14.5 | 11.5 | 11.6 |
| 95 | 50 | 12.0 | 2.56 | 8.0 | 13.3 | 5.35 | 5.6 | 14.6 | 11.4 | 11.7 |
| 95 | 10 | 11.6 | 2.10 | 6.2 | 14.0 | 4.45 | 5.1 | 14.6 | 12.6 | 10.6 |
| **AVG** | | **11.9** | | **8.2** | **13.8** | | **9.2** | **14.4** | | **11.5** |
| **STD** | | **0.2** | | **2.1** | **0.3** | | **2.7** | **0.2** | | **3.3** |

Table 5. Fin surface peak frequency ($f_{pk}$), the peak level and peak quality factor ($Q$) at 21 locations.

Since the fin was excited by coherent structures shed from the cylinder, it is of interest to examine the coherence of these fin surface motions. The coherence can be quantified as

$$\gamma_{xy}^2 = \frac{|G_{xy}(f)|^2}{G_{xx}(f)G_{yy}(f)}$$

(Bendat & Piersol, 1980), where $G$ is either the cross spectra ($G_{xy}$) or autospectra ($G_{xx}$, $G_{yy}$) at a given frequency ($f$) between given points ($x,y$) on the fin surface. Figure 14 shows the coherence as a function of separation distance between each of the measurement points and the reference point (80% span, 90% chord). As expected the coherence decreases with increasing separation distance and frequency. The shedding frequency at this speed ($U_{sp}$ = 3.62 m/s) was 14.5 Hz, which has high coherence over nearly the entire range. The high coherence is reflective of the fact that the fin vibration is dominated by a single mode.

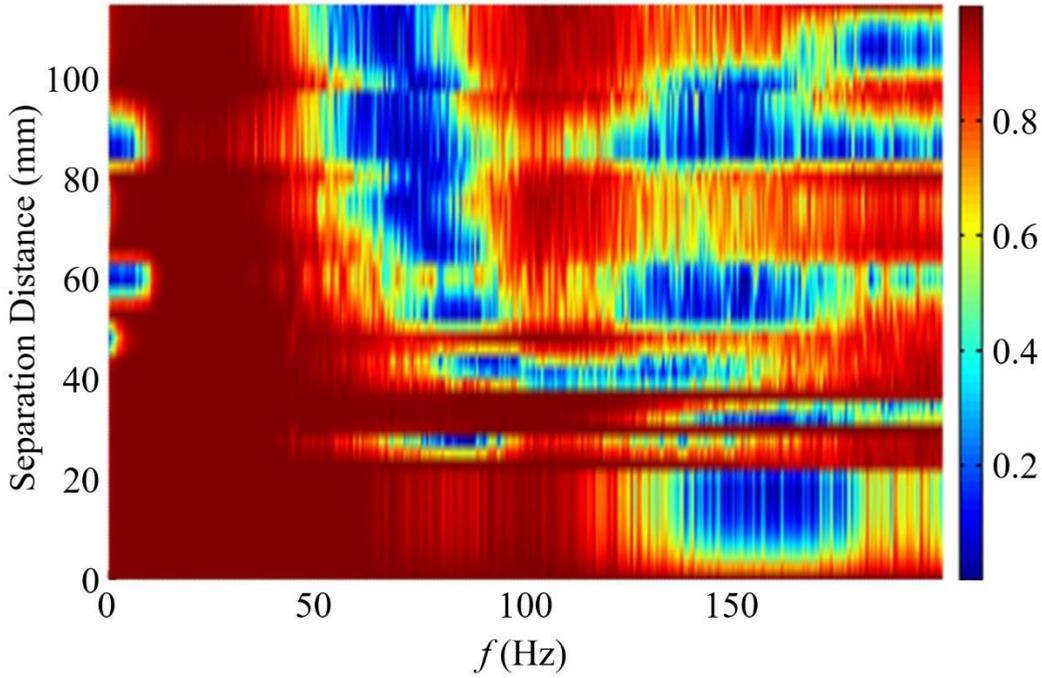

Figure 14. Contour plot of the blade vibration coherence as a function of frequency and separation distance relative to the 80% span and 90% chord position ($U_{sp}$ = 3.62 m/s).

## *5.3 Fin constraint loading*

The load cell measured the constraint lift ($\mathcal{L}$), drag ($\mathcal{D}$), torque ($T_q$) and drag moment ($M_D$), which is the constraint moment about the axis perpendicular to the chord length. Time traces show the shedding frequency dominates the oscillations with additional broadband content, and $\mathcal{L}$ was 180° out-of-phase with the other constraint loads. The average constraint forces were scaled using the traditional scaling for the coefficients of lift $\left(C_L \equiv \mathcal{L}/(0.5\rho U_{sp}^2 bc)\right)$ and drag $\left(C_D \equiv \mathcal{D}/(0.5\rho U_{sp}^2 bc)\right)$. The moments were similarly scaled though with the span squared to account for the moment arm, $C_{MD} \equiv M_D/(0.5\rho U_{sp}^2 b^2 c)$ and $C_{Tq} \equiv T_q/(0.5\rho U_{sp}^2 b^2 c)$. The scaled mean constraint loading versus $Re_c$ are provided in Figure 15, which shows that they are nearly independent of Reynolds number over the range tested ($10^5$ to $1.8\times10^5$). The average values are -0.200 ± 0.018, 0.114 ± 0.007, 0.040 ± 0.002 and 0.0287 ± 0.002 for the $C_L$, $C_D$, $C_{MD}$ amd $C_{Tq}$, respectively. Here the uncertainties are twice the standard deviation from measurements at different Reynolds numbers. Note that the mean negative lift force is consistent with non-zero mean fin tip deflections at a given speed.

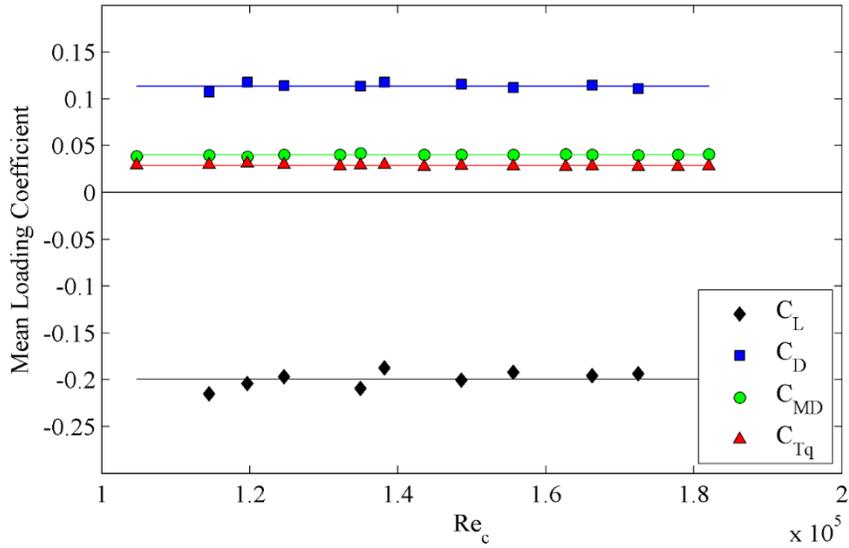

Figure 15. Scaled mean constraint forces and moments plotted versus the chord based Reynolds number ($Re_c$). The solid lines correspond to the average value over all the speeds tested.

Measurement of the torque and drag-moment enables the calculation of the center of pressure in two directions. The torque was induced by the lift and drag forces, which suggests that the resultant load acts a distance $R_T$ from the center of the load cell. This distance was determined by dividing the torque by the magnitude of the lift and drag forces, and should be a constant since the constraint loading coefficients were constant and the span was held constant. The resultant torque moment arm was $R_T = 29.2 \pm 2.0$ mm, where the uncertainty is twice the standard deviation from results at each test speed. This shows that the center of pressure was located more than half a chord length from the load cell center due to the forward sweep of the fin. In addition, the drag moment can be used to determine the spanwise location of the center of pressure. Since the load cell was zeroed before testing, the weight was assumed negligible, which makes the force parallel to the chord the only loading considered. The resulting spanwise center of pressure was at $z = 119 \pm 14$ mm, where once again the uncertainty is twice the standard deviation from results at each test speed. This shows that the center of pressure was slightly above the center of the span and experienced larger variations than the torque.

Since the constraint loading time traces showed broadband content, it is useful to examine the constraint loading fluctuations in more detail. The RMS constraint forces and moments were scaled with their respective mean values and plotted versus the chord based Reynolds number in Figure 16. Linear regression analysis of these curves show that the slope of the fluctuating drag and torque are statistically significant (p-value < 0.05). The resulting linear fits for the fluctuating drag and torque curves are $\mathcal{D}'/|\mathcal{D}| = -1.07 \times 10^{-6} Re_c + 0.405$ and $T_q'/|T_q| = -3.92 \times 10^{-6} Re_c + 2.21$, respectively. Conversely, the fluctuating lift and drag-moment curves are independent of Reynolds with $\mathcal{L}'/|\mathcal{L}| = 0.326 \pm 0.038$ and $M_D'/|M_D| = 1.935 \pm 0.174$, where the uncertainty is twice the standard deviation. In general, these results are more scattered than the mean due to the broadband nature of the cylinder vortex shedding coupled with the structural response of the fin. Consequently, it is more informative to investigate the spectral content of these fluctuations.

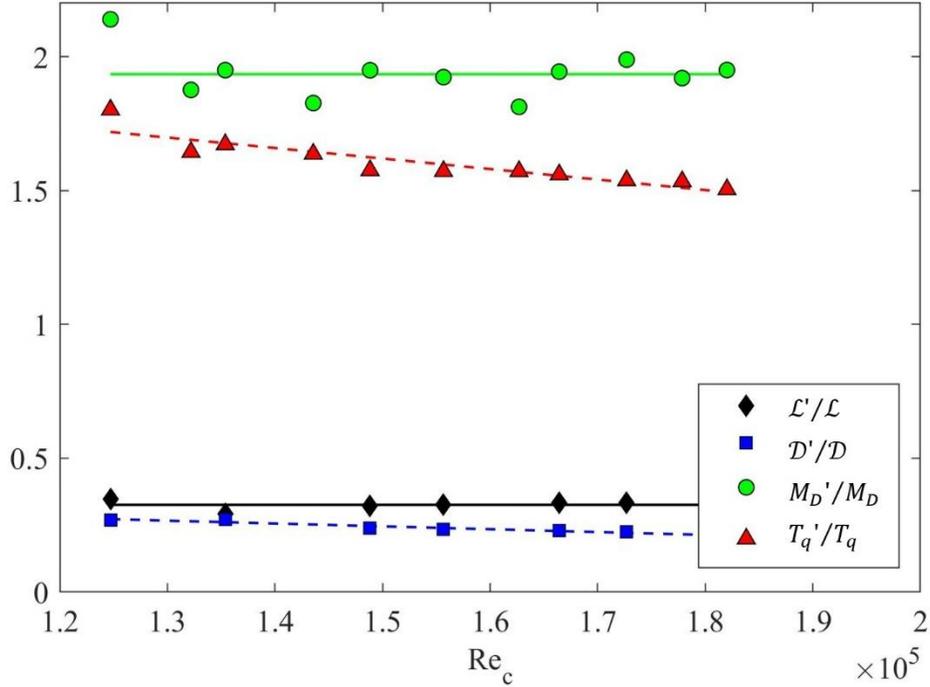

Figure 16. The RMS constraint forces and moments scaled with their respective mean values and plotted versus the chord based Reynolds number. The solid lines correspond to average values and dashed lines are linear fits.

The power spectral density of the lift constraint force was examined to assess the repeatability as well as sensitivity to angle-of-attack. Figure 17a shows repeated (7 total) measurements of the power spectral density of the fin constraint lift force ($\Phi_\mathcal{L}$) at $\alpha = -0.8°$ and $U_{sp} = 3.1$ m/s. Each of the tests were performed with the tunnel starting at the no-flow condition, ramping to the desired speed and then acquiring steady state data. These results consistently show a peak RMS lift force at the shedding frequency of $16.66 \pm 0.38$ N$_{RMS}$, or approximately $\pm 2.3\%$ uncertainty. The peak (shedding) frequency varied by $\pm 1.5\%$ at $13.67 \pm 0.02$ Hz. Figure 17b compares the sensitivity of the lift power spectral density to angle-of-attack at the same speed ($U_{sp} = 3.1$ m/s). The angle-of-attack appears to have minimal impact on the fluctuating fin lift forces at the shedding frequency, with the peak RMS lift force ($16.69 \pm 0.43$ N$_{RMS}$) within the uncertainty range of the repeated $-0.8°$ condition. This does not imply that angle-of-attack does not impact the mean lift force, only the fluctuating lift force. Note that due to concerns of overstraining the load cell and/or exceed the designed fin deflections during these tests, a mechanical stop that bracketed the fin tip was installed only for these measurements. Figure 17b includes a repeated condition with the mechanical stop removed that shows that the influence of this stop had negligible impact on the fin loading. Overall these repeatability results provide confidence in both the repeatability of the facility and the fluid loading.

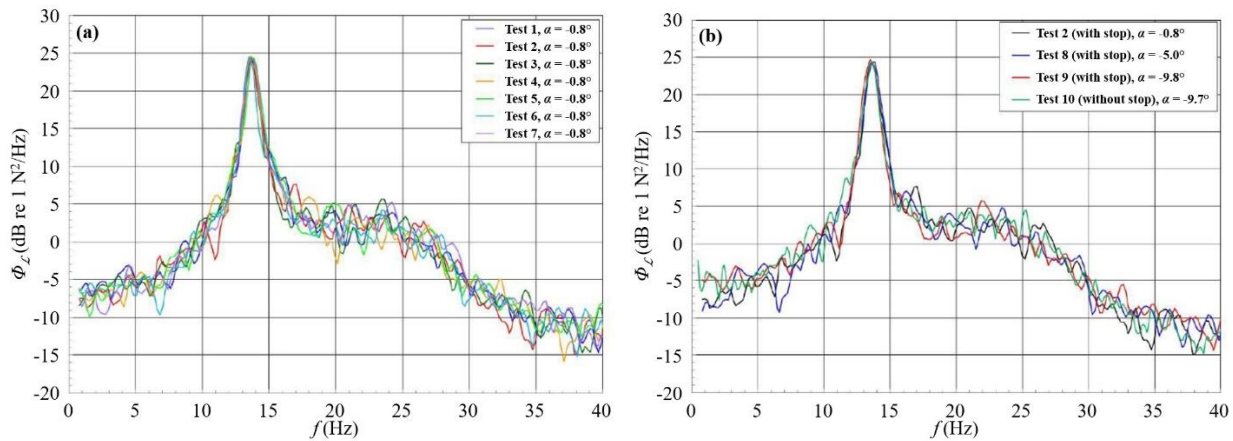

Figure 17.(a) Power spectral density of the fin constraint lift force at $U_{sp}$ = 3.1 m/s and a fixed angle-of-attack (-0.8°) to demonstrate repeatability. (b) Lift force power spectral density of at $U_{sp}$ = 3.1 m/s with varying angle-of-attack.

The power spectral density of the constraint loading components at $U_{sp}$ = 3.44 m/s and an angle-of-attack of -9.7° is provided in Figure 18a. These results show that the fin lift force at the shedding frequency of 14.9 Hz was ~10 dB greater than the drag component. The drag force and drag moment have an additional peak at ~44 Hz, which was not observed in the torque or lift force. This is due to the first streamwise bending mode (Table 2), which since it is a streamwise mode it makes sense that it was only observed in the loading associated with the drag. The first cross-stream bending mode (~23 Hz) was not apparent in the results, but there is a slight shelf coming down from the shedding frequency that holds until ~23 Hz before dropping more rapidly. At the shedding frequency, the RMS fin torque (twist) was ~0.275 Nm (-11.2 dB), which is nearly an order of magnitude less than the moment applied during the static load testing. This is consistent with the RMS fin tip twist measurements that showed only a few tenths of a degree movement. The speed dependence of these results were also examined with the results provided in Figure 18b. Here the fin lift constraint power spectral density spanning the range of speeds tested are provided. It is apparent that the peak frequency and amplitude increase with increasing speed. Note that the lowest speeds appear to have a weak secondary peak forming between 20 and 25 Hz, which corresponds to the first cross-stream bending mode (Table 2). The peak frequencies match the cylinder shedding frequency with a constant Strouhal number of 0.263 ± 0.007, where the uncertainty is twice the standard deviation. The peak amplitudes were 11.2, 13.4, 15.3, 16.7 21.7 and 23.1 N for $U_{sp}$ = 2.5, 2.7, 2.95, 3.05, 3.3 and 3.4 m/s, respectively. These amplitudes are well approximated as a power-law function with the amplitude being proportional to $U_{sp}^{2.6}$.

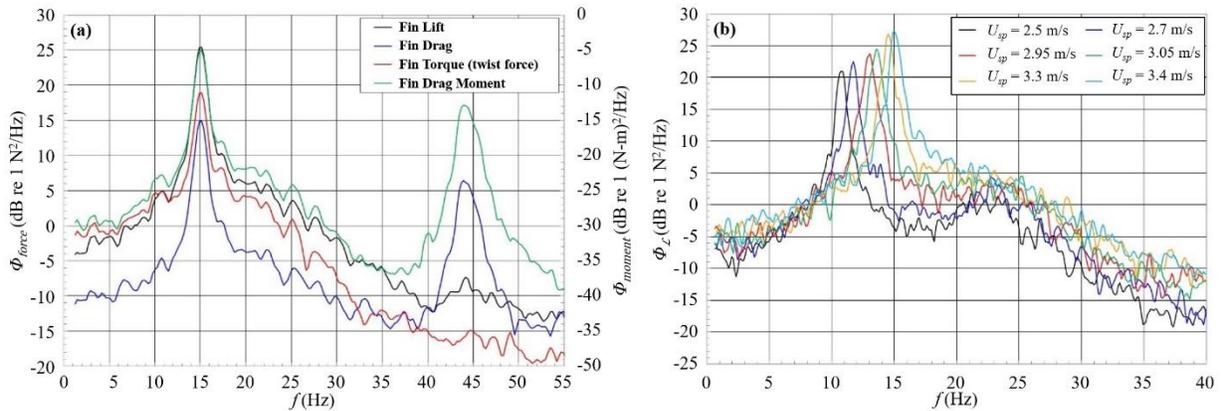

Figure 18. (a) Power spectral densities for the fin constraint loading from a single test condition ($U_{sp}$ = 3.44 m/s). (b) Fin lift constraint force power spectral density measured over a range of tunnel speeds.

# 6  Summary and Conclusions

The current study presents experimental results from a high-amplitude, low-frequency flow-induced fin oscillation, which can be used for development and validation of FSI models. The experiment was conducted in a recirculating water tunnel at speeds primarily between 2.5 and 3.6 m/s. A 60 mm diameter cylinder with a roughened leading edge was mounted near the test section inlet and a backward facing (i.e. hydrodynamic leading edge facing downstream) swept fin located directly downstream of the cylinder. The fin had a 230 mm span, 50 mm chord that was necked down to 35 mm at the 25% span location and positioned at a -9.64° angle-of-attack. The fin was mounted on a load cell that must be considered an integral part of the test model because a significant amount of the fin movement was generated due to the load cell. The structural response of the fin and load cell were characterized *in situ* via in-air and in-water modal impact tests as well as static loading tests. The flow-field and structural response were characterized with measurements of the cylinder base and dynamic pressure, wake flow-field (high-speed PIV and LDV), fin motion (high-speed imaging of the tip and laser vibrometer on the surface) and constraint loading (lift, drag, torque and drag moment).

The geometry (tunnel, cylinder, fin and load cell), fluid properties and material properties/response are provided in the experimental methods (§2). The inlet flow was also required for FSI modelers, but the experimental setup prevented direct measurement of the velocity upstream of the cylinder. Consequently, the inlet flow was characterized via velocity measurements between the cylinder and fin, including local free-stream speed, centerline velocity and scaled streamwise velocity profiles spanning the cylinder wake. In addition, the cylinder base pressure and shedding frequencies are provided for the smooth and rough ($St = 0.26$) configurations.

Given the inlet flow and the resulting cylinder shedding, the force fin behavior was characterized with measurements of the flow-field upstream of the fin, fin motion (tip and surface) and the constraint loading (lift force, drag force, drag-moment and torque). The mean results showed that the cylinder wake with and without the fin installed were consistent with canonical cylinder wake studies. The mean fin tip motion showed minimal twist (≤ 0.3°) for all test conditions, while the tip deflections linearly increased with the chord-based Reynolds number (though a power-law fit produced equally valid results for the given range). The peak-to-peak fin tip deflections were up to 5.8 times the mean deflection at a given speed. These mean deflections produced constant mean scaled constraint loads over the range tested ($10^5 \leq Re_c \leq 1.8 \times 10^5$), which suggests that the constraint loading was proportional to the square of the tunnel speed. Combining the mean force and moment measurements allowed an estimate of the center of pressure for the resultant load. Independent of the tunnel speed at the test angle-of-attack (-9.7°), the torque moment arm was 29.2 mm from the center of the load cell and the resultant center of pressure was located at $z = 119$ mm.

While the mean results show consistent behavior, the fluctuating cylinder wake, fin motion and fin loading are more revealing of the underlying physical processes. The cylinder wake showed that the fluctuating wake profiles were also consistent with traditional wake behavior. Furthermore, wavenumber spectral analysis showed that the wake was unaltered upstream of the fin, which suggests that there was no feedback mechanism between the fin and the upstream cylinder wake. Repeated measurements of the power spectral density of the fin loading demonstrated the repeatability of the experiment. In addition, variation of the angle-of-attack showed that the power spectral density distribution was insensitive to angle-of-attack, even though the mean loading was dependent on the angle-of-attack. RMS fin tip motions showed negligible fluctuations in twist (~0.3°) and the dominate tip deflection frequency coincided with the cylinder shedding frequency. The RMS constraint loading scaled with the corresponding mean value showed that the fluctuating lift force and drag-moment were constant, but the drag force and torque were linearly proportional to the chord based Reynolds number.

The fin surface deflections and constraint loading had a higher temporal resolution, which enabled more in-depth analysis of the broadband content of the fin motion and loading. The fin surface deflections show three distinct peaks that match the cylinder shedding frequency ($St = 0.26$), the first cross-stream bending mode (23 Hz) and the second cross-stream bending mode (103 Hz). The cylinder shedding frequency also dominates all the constraint loading measurements with the increase in amplitude proportional to the tunnel speed raised to the power of 2.6. The constraint loading spectra also shows a weak peak at the first cross-stream bending mode, which was more prevalent at lower speeds. Conversely, the first streamwise bending mode (44 Hz) that was not observed in the fin surface deflections produced significant peaks in the drag force and drag moment spectra. Both of these observations are consistent given that the laser vibrometer used for the fin motion was only sensitive to motion perpendicular to the surface (i.e. primarily the cross-stream direction) while the drag and drag moment were more sensitive to streamwise oscillations. Hence the first streamwise bending mode was only weakly observed in the lift force spectra.

This paper provides a unique flow that will challenge FSI modeling capabilities due to the relatively large fin oscillations induced by a complex flow-field. Presented is the information required to setup the initial conditions for modelers as well as several independent measurements characterizing the response of the forced fin. This rich dataset will give modelers the ability to identify strengths and weaknesses of various modeling approaches. This will become more important as composite materials are explored for propulsion schemes, especially for marine applications.

## Acknowledgements

The authors would like to thank the technical staff at the Applied Research Laboratory, especially Neil Kimerer assistance in the load cell design. We would also like to thank Dr. Peter Chang and Dr. Scott Black for guidance throughout planning and execution of the experiment. This research was sponsored by DARPA under Contract N0002402-D-6604 (Dr. Christopher Warren, Program Manager). The content of this document does not necessarily reflect the position or the policy of the US Government, and no official endorsement should be inferred.